\documentclass{aa} 
 
\usepackage{graphicx} 
\usepackage{epsf} 
 
\newcommand{\Lstar}{\mbox{$L_*$}} 
\newcommand{\Lbol}{\mbox{$L_{\rm bol}$}} 
\newcommand{\Lsun}{\mbox{$L_\odot$}}

\newcommand{\cps}{\mbox{cnt~s$^{-1}$}}

\newcommand{\msun}{\mbox{$M_\odot$}} 
\newcommand{\mdot}{\mbox{$\dot{M}$}} 
\newcommand{\Rstar}{\mbox{$R_*$}}

\newcommand{\vinf}{\mbox{$v_\infty$}}

\newcommand{\be}{\begin{equation}} 
\newcommand{\ee}{\end{equation}}

\newcommand{\Lx}{\mbox{$L_{\rm X}$}} 
\newcommand{\Tx}{\mbox{$T_{\rm X}$}}

\newcommand{\xmm}{\mbox{{\sc XMM}-{\it Newton}}} 
 
\def \etal\,{et~al.\/} 
\def\lesssim{\mathrel{\hbox{\rlap{\hbox{\lower4pt\hbox{$\sim$}}}\hbox{$<$}}}} 
\def\gtrsim{\mathrel{\hbox{\rlap{\hbox{\lower4pt\hbox{$\sim$}}}\hbox{$>$}}}}

\def\figurenum#1{\def\thefigure{#1}\let\@currentlabel\thefigure 
\addtocounter{figure}{\m@ne}} 
\def\figcaption{\@ifnextchar[{\@xfigcaption}{\@figcaption}} 
\def\@figcaption#1{{\def\@captype{figure}\caption{#1}}} 
 
\def\@xfigcaption[#1]#2{{\def\@captype{figure}\caption{#2}}} 
\def\fnum@figure{{\rm Figure\space\thefigure:}} 
\def\fps@figure{bp} 
\def\eps@scaling{.95}\def\epsscale#1{\gdef\eps@scaling{#1}} 
\def\plotone#1{\centering \leavevmode 
\epsfxsize=\eps@scaling\columnwidth \epsfbox{#1}}

\begin{document} 
 
\title{The Conspicuous Absence of X-ray Emission from  
Carbon-Enriched Wolf-Rayet Stars} 
 
\author{L.~M.~Oskinova\inst{1}, R.~Ignace\inst{2}, W.-R.~Hamann\inst{1}, 
A.\,M.\,T.~Pollock\inst{3}, J.\,C.\,Brown\inst{4}} 
\institute{Astrophysik, Univerit{\" a}t Potsdam, Am Neuen Palais 10, D-14469 Potsdam, Germany 
\and 
Department of Astronomy, University of Wisconsin, Madison, WI 53706, USA 
\and 
ESA/Vilspa, Apartado 50727, 28080 Madrid, Spain  
\and 
Department of Physics and Astronomy, University of Glasgow, Glasgow, G12 8QQ,  
 UK} 
 
\offprints{lida@astro.physik.uni-potsdam.de} 
\date{Received <date>; Accepted <date>} 
\authorrunning{Oskinova \etal\,} 
\titlerunning{Conspicuous absence of X-rays from WC stars} 
 
\abstract{The carbon-rich WC5\,star WR\,114 was not detected during a 
15.9~ksec \xmm\, observation, implying an upper limit to the X-ray  
luminosity of \Lx\,$\lesssim\,2.5\times 10^{30}$\,erg\,s$^{-1}$ and  
to the X-ray to bolometric luminosity ratio of  
\Lx\,/\Lbol\,$\lesssim\,4\times10^{-9}$. This confirms 
indications from earlier less sensitive measurements that there 
has been no convincing X-ray detection of any single WC~star. This 
lack of detections is reinforced by \xmm\, and {\sc Chandra} 
observations of WC~stars. Thus the conclusion has to be drawn that  
the stars with radiatively-driven stellar winds of this particular  
class are insignificant X-ray sources. We attribute this  
to photoelectronic absorption by the stellar wind. The high 
opacity of the metal-rich and dense winds from WC stars puts the radius 
of optical depth unity at hundreds or thousands of stellar radii for 
much of the X-ray band. We believe that the  
essential absence of hot plasma so far out in the wind exacerbated by the 
large distances and correspondingly high ISM column densities  
makes the WC stars too faint to be detectable with current technology.  
The result also applies to many WC~stars in binary systems, 
of which only about 20\% are identified X-ray  
sources, presumably due to colliding winds. 
 
\keywords{Stars: individual: WR\,114 (HD~196010) -- stars: winds, 
 outflows -- stars: early-type -- X-rays: stars}  
}  
 
\maketitle 
 
\section{Introduction} 
 
The few major classes of star whose spectral properties are dominated 
by their mass loss in the form of stellar winds include the hot 
OB~stars, with roughly solar chemical composition and mass-loss rates 
\mdot$_{\rm OB}\approx 10^{-7}$\msun\,yr$^{-1}$ and the Wolf-Rayet (WR) 
stars with $\mdot \approx 10^{-5}$\msun\,${\rm yr}^{-1}$.  The WR stars 
come in two main types:  WN~stars with enriched helium content and 
deficient hydrogen and WC~stars with no hydrogen but strongly enhanced 
carbon and oxygen.  It is thought that O~stars with sufficient initial 
masses evolve to WR~stars, passing through the WN~star stage and 
become WC~stars whose own likely end is in gravitational  
collapse (Woosley \etal\, 2002).  
 
\begin{figure*}[hbtp] 
  \centering 
  \mbox{\epsfxsize=.8\textwidth \epsffile{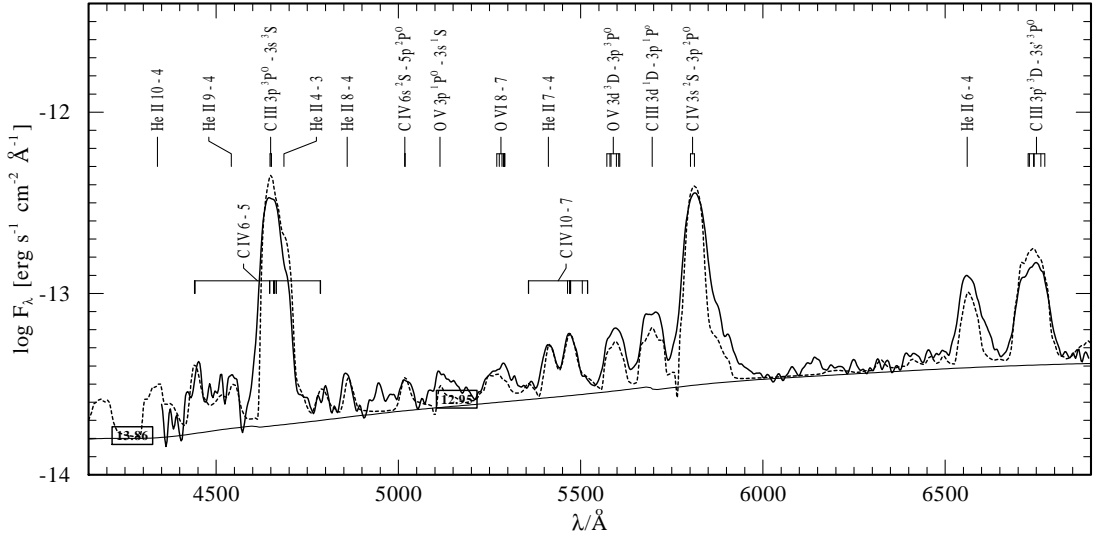}} 
 \caption [ ]{ 
Observed visual spectrum (solid line) of WR\,114 from 
the atlas of Torres \& Massey (1987), compared to a 
synthetic model (dotted line) calculated with the Potsdam non-LTE 
stellar wind code (Gr{\" a}fener \etal\, 2002). The thin smooth line is  
the model continuum. The model parameters are: effective temperature  
$T_*$= 80\,kK, luminosity log $L_*/$\Lsun = 5.2, mass loss rate  
$\log \mdot = -5.0$\msun\,yr$^{-1}$, terminal wind 
velocity \vinf\, = 2000 km/s, chemical composition He:C:O = 55:40:05, and a 
clumping contrast of $D=10$. The calibration of the observation is 
confirmed by $v,b$ photometry (small boxes). The synthetic spectrum is 
scaled with the distance modulus of WR\,114 (11.5\,mag from the 
membership in the Sgr\,OB1 association) and reddened with $E_{b-v} = 
1.2$\,mag (corresponding to $E_{B-V}$=1.45\,mag) in order to reproduce the 
observed slope.} 
\label{fig:f3}  
\end{figure*} 
 
The mass loss from OB~stars is reasonably well explained in the  
framework of the CAK theory (Castor, Abbott \& Klein 1975) 
of line-driven stellar  
winds. There has also been progress in the understanding  of  
winds from WR stars. Recent advances in atmospheric modeling  
demonstrate that the mass loss of WC~stars, like that of OB and WN stars,  
is also generally explicable by radiative driving 
(e.g. Gr{\"a}fener \etal\, 2002). 
 
However, the X-ray emission of OB and WR~stars remains enigmatic. 
Discovered with the {\sc Einstein} observatory (Seward \etal\, 1979;  
Harnden \etal\, 1979), soft X-ray emission from hot stars  was  
initially attributed to shocks due to the intrinsic instability  
of the radiative driving mechanism (Lucy \& White 1980; Lucy 1982). 
Further developments of the theory by Feldmeier \etal\, (1997a,b) 
were reasonably able to reproduce the low-resolution spectra of O~stars  
observed by the {\sc Rosat} observatory but are now facing new 
challenges with the recent advent of high-resolution spectroscopic data 
from the \xmm\ and Chandra grating spectrometers.  
 
Application of the theory to X-ray production in the high-density winds 
of WR~stars has yet to be done in detail. Importantly, however, Gayley 
\& Owocki (1995) have considered the effects of multiple scattering in 
WR~winds for the wind instability mechanism. They found that multiple 
scattering has a suppressing effect on the intrinsic instability that is 
normally thought to explain the X-ray production of the less dense 
winds of OB~stars. However, a significant residual instability remains 
and is sufficient to produce extensive structure in the wind. 
Theoretically, this process may lead to shock formation and 
plasma heated to X-ray emitting temperatures of few million Kelvin. 
Arguably, the presence of lines of highly ionized species such as 
O{\sc vi} in the UV spectra of WR~stars indicates the presence of 
X-ray photons in the line-forming regions of the wind (e.g. Gr{\" 
a}fener \etal\, 2002). 
 
The study of X-rays from single WR~stars and their spectral properties 
provides much needed new insight into the understanding of stellar 
winds.  WR~stars have short lifetimes and, although evolutionarily linked 
to O~stars and thus located in the same OB~stellar associations, 
are relatively rare and distant. Most WR stars are X-ray faint 
and thus hard to detect. Among the exceptions are the binary systems, 
where stronger and harder X-ray emission due to wind-wind collisions 
suggests criteria for distinguishing single and double stars. 
Unfortunately, little is learned about the X-rays in the WR~wind itself 
from such systems. 
 
In this paper we concentrate on the study of single WC stars. 
On average, the WC stars, with absolute visual magnitudes not 
less then $M_{\rm v}\approx -4.5$\,mag (van\,der\,Hucht 2001) 
are less luminous than the O and WN stars from which they evolve 
because of the reduction in mass  during evolution and the very high  
effective temperatures of up to $10^5$\,K that shift the energy distribution 
towards the EUV.  
For comparison, the visual absolute magnitudes of O~type stars lie between   
$M_{\rm v}\approx\,-4.5~{\rm and}~$-7.5\,mag (Vacca \etal\, 1996).  
 
It was clear from the first reports (Pollock 1987b) that WC~stars  
are generally fainter in X-rays than other hot stars, the majority 
of measurements yielding merely upper limits.  
We have now taken advantage of the unprecedented sensitivity of \xmm\,  
to observe the carbon type WC5 star, WR\,114 (\,=\,HD\,169010) 
for an exposure time, after excluding bad time intervals,   
of 15.9\,ksec. To our knowledge, this is the first such sensitive  
exposure of a single WC~star. Nevertheless, \xmm\, was unable to obtain  
even a 2$\sigma$ detection of WR~114 in any energy  
band. In \S 2 we summarize the stellar parameters of WR~114 and in 
\S 3, we describe the observations of WR~114 with \xmm . In \S 4  
we review all currently available X-ray  
observations of single WC~stars and in \S 5 we briefly review X-rays  
observations from binary WC~stars. The results are discussed in \S 6  
and conclusions are drawn in \S 7. 
 
\begin{figure*}[hbtp] 
  \centering 
  \mbox{\epsfxsize=.8\textwidth \epsffile{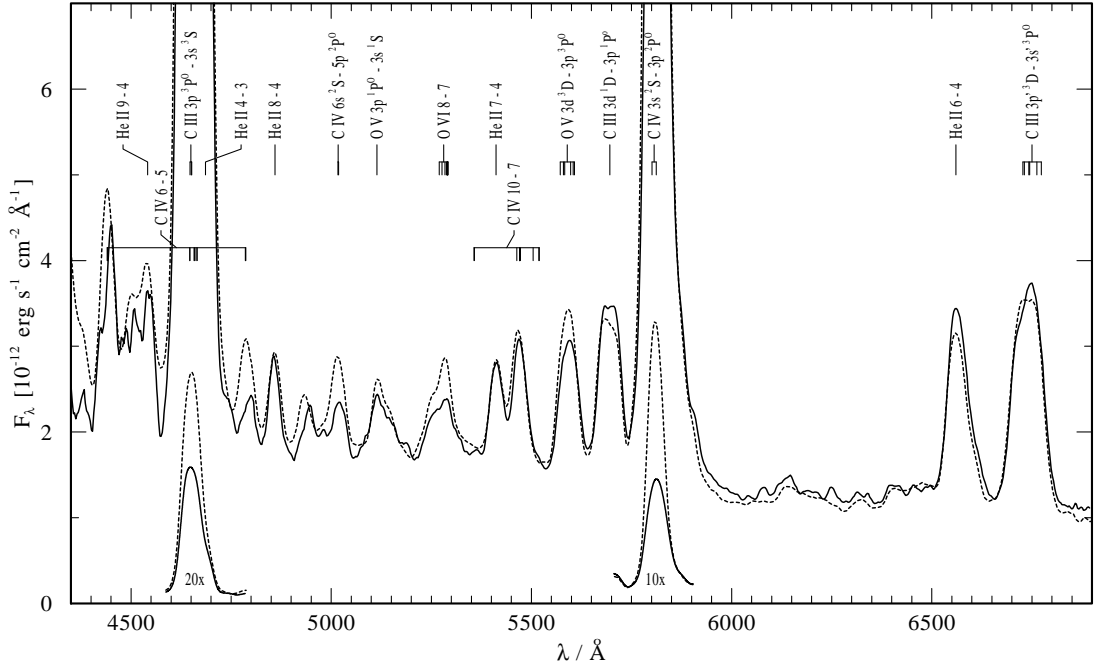}} 
 \caption [ ]{ 
Observed visual spectrum of WR\,114 (solid line) from 
the atlas of Torres \& Massey (1987), compared to 
an observation of the single ``standard'' WC5 star WR\,111 (dashed line). 
For this comparison, the WR\,114 spectrum has been dereddened with 
$E_{b-v}=1.0$\,mag in order to compensate for the stronger color excess 
of that star, and scaled by a factor of 3.0 to the flux level of  
WR\,111. Obviously, the spectra of both stars are very similar and  
indicate similar stellar parameters. Therefore we see no reason to  
suspect that the WR\,114 spectrum is a binary composite, in contrast  
to the ``diluted emission line'' entry given in the  
van\,der\,Hucht (2001) catalogue.} 
\label{fig:f4}  
\end{figure*} 
 
\section{Stellar parameters of  WR\,114} 
 
WR\,114 (WC5) is an ordinary representative of its spectral type. 
Its optical spectrum, the defining characteristic of the WC type,  
is well reproduced by the Potsdam standard atmospheric code,  
the latest version of which (Gr{\"a}fener \etal\, 2002) we employed  
to infer the stellar parameters. The best fit to the observed spectrum, 
shown in Fig.\,1, was obtained with an effective temperature  
$T_*$= 80\,kK, luminosity $\log\Lbol/$\Lsun= 5.2, mass loss rate 
$\log\mdot = -5.0\,\msun\,{\rm yr}^{-1}$, terminal wind velocity  
\vinf\, = 2000\,km s$^{-1}$, chemical composition He:C:O = 55:40:5 by mass,  
and a clumping contrast of $D=10$ (for definition of the latter see  
Hamann \& Koesterke 1998). 
 
In ``The {\sc vii}$^{th}$ Catalogue of Galactic WR Stars''  
(van\,der\,Hucht 2001) a number of galactic WC~stars, including WR~114,   
are considered to have an OB companion because of the weakness of  
their emission lines in comparison with other stars of the same subtype.   
However, this ``diluted emission line'' or d.e.l. evidence is not very  
strong, as WR spectra depend on {\em two} parameters -- effective temperature  
and mass-loss rate -- and it is not clear {\it a priori} whether there  
exists a variety of line strengths within one spectral subclass. 
In Fig.\,2 
we compare the spectrum of WR\,114 with the prototype apparently single WC5 
WR\,111, which was also one of the stars employed by van der 
Hucht (2001) for his comparison. The plot reveals that most spectral 
lines are very similar or even identical in strength and profile. Dilution 
by a companion's continuum would affect {\em all} lines 
uniformly. Therefore we doubt the ``diluted emission line'' case argued 
given by van der Hucht (2001) for the binary status of  WR\,114. 
 
WR~114 was observed but detected neither in the mid-infrared  
(Smith \& Houck 2001) nor as a source of non-thermal radio emission  
(Chapman \etal\, 1999), either of which would have provided 
supporting evidence for binarity.
 
\section{XMM-Newton Observations of WR\,114} 
 
\xmm\, observed WR\,114 on 23~March~2002, as summarized in Table~1,  
for about 23.4\,ksec in total. Two exposures were taken of 
16.9\,ksec and 6.5\,ksec although the  
latter was affected by a spacecraft pointing error and is 
unusable. 
 
The observatory is described in a special letters issue  
of {\it Astronomy \& Astrophysics}~(Vol.~365, L1). Relevant data were 
obtained with the European Photon Imaging Camera (EPIC) and the 
Optical/UV Monitor (OM). EPIC is equipped with two identical MOS~cameras  
and a pn~camera.  These provide CCD imaging spectroscopy with energy  
coverage between $0.2-15$\,keV in a 30\,arcmin diameter field-of-view (FoV)  
with an angular resolution of approximately 12\,arcsec~FWHM at 1.5 keV.  
Observations were made using the full-window mode and the medium optical  
blocking filter. OM provides photometric coverage between 1700\,\AA\, and  
6500\,\AA\, of the central 17\,arcmin region of the X-ray FoV, permitting  
simultaneous observations in both the X-ray and the UV/optical bands. 
 
Preliminary examination of the standard pipeline processed images 
of the full 16.9\,ksec exposure immediately gave a puzzling result. 
Even though the star is obvious in the OM optical V~band image at an 
offset of only $0\farcs02$  from its expected position, 
as shown in Fig.~3, it is nowhere to be seen in any of the 
X-ray images such as the one shown in Fig.~4. 
The coordinates of the star and instrument are shown in 
Table~\ref{tab:tab1} with other relevant data. We then performed a 
thorough data reduction using the \xmm\, Science Analysis 
System software (SAS v\,5.3.3). The latest calibration files were 
used and data were filtered to select good event patterns ensuring 
that only X-ray related events were considered. After exclusion of 
contaminated time intervals due chiefly to soft proton flares, the 
useful exposure time dropped from $16.9$\,ksec to  $15.9$\,ksec for 
each of the MOS cameras and $10.2$\,ksec for pn. The extracted  
images were carefully investigated. Some X-ray sources are obvious  
(see Fig.~4), but the area in the vicinity of the optical image of  
WR\,114 contains no X-ray source in either MOS1, MOS2 or pn, individually  
or in combination. 
Exposure times, corrected for vignetting, were then determined for  
this area.  Since there are no nearby X-ray or bright UV sources which 
could contaminate the image, and as the number of counts per pixel in 
the area is the order of unity, Poisson statistics are applied. 
 
\newcommand{\CoCo}[8]{$#1^{\rm h}#2^{\rm m}#3\fs#4~#5\degr#6\arcmin#7\farcs#8$} 
\begin{table} 
\begin{center}
\caption {\xmm\, Observations of WR\,114$^a$ {\label{tab:tab1}}}
\begin{tabular}{l l}\hline\hline 
\rule[0mm]{0mm}{4.0mm} 
 ~Parameter                       & ~Value                       \\  
\hline 
 Start (UT)                       & 2002-03-23 13:46                      \\  
 End (UT)                         & 2002-03-23 18:44                      \\  
 Exposure [ksec]                  & 15.9~(MOS1,2),~~10.2~(pn)             \\ 
 Pointing coordinates             & \CoCo{18}{23}{16}{39}{-13}{43}{25}{8} \\ 
 WR114 coordinates                & \CoCo{18}{23}{16}{30}{-13}{43}{26}{0} \\  
 Count Rate[$10^{-4} $\,\cps]     & $<2.6$~(MOS1,2),~~~$<6.7$~(pn)        \\
 Flux[erg\,cm$^{-2}$\,s$^{-1}$]   & $<5.3\times 10^{-15}$                 \\   
 Luminosity\,[erg\,s$^{-1}$]  & $<2.5\times 10^{30}$                  \\ 
\hline\hline 
\end{tabular} 
\end{center} 
{\small $^a$ One 6.4\,ksec exposures was 
          of no     scientific use since 
         the manoeuvre to the target failed. The count rates are 
         1$\sigma$ vignetting-corrected source upper limits. The 
         flux and luminosity are for the MOS1 upper limit in the  
         0.2-10 keV range and have been corrected for interstellar  
         absorption of $N_{\rm H}\approx 5.5\times 10^{21}{\rm cm}^{-2}$.}  \\
 
\end{table} 
 
We used the SAS's source detection task {\sc edetect\_chain}  
with the recommended $4\sigma$ detection likelihood threshold of 10. 
Attempts to detect the source were made both in 
narrow band and broad band images without success. Neither was WR\,114  
detected with thresholds as low as 2$\sigma$.  
Thus the conclusion is unavoidable that WR\,114 was not detected by our  
\xmm\, pointed observation.  
 
We are able to set an upper limit on the X-ray count rate of 
WR\,114. The on-axis 1.5\,keV point spread functions (PSFs) of the 
X-ray telescopes have FWHM values of about 4.4\,arcsec for MOS1,2 and 
$<12$\,arcsec for pn. The EPIC source positions, 
even for faint sources close to the detection limit, have typical 90\% 
confidence radii of about 2-5\,arcsec, limited by the statistical 
accuracy of the measurements (Watson \etal\, 2001).  Finally, one pixel 
in the EPIC~CCDs covers about 1.1~arcsec for the MOS cameras and 4.1~arcsec 
for the pn.  We chose to take a 3$\times$3 pixel box covering  $15\times 
15$\,arcsec centered on the OM position of 
WR\,114. Using the SAS task {\sc esenmap},  
we estimated the 1$\sigma$ count rate upper limits 
reported in Table~\ref{tab:tab1}.  
Following the concept of \xmm\, for which each of the three detectors 
(MOS1, MOS2 and pn) is located in the focal plane of its own X-ray 
telescope and observations are made simultaneously, we have to accept 
as an upper limit on the source the lowest count rate of all three 
detectors, namely MOS1 (see Table~\ref{tab:tab1}). 
 
Of crucial influence to the X-ray luminosity determination is the 
interstellar absorption.  Because no UV observation of WR\,114 
is available, we used the visual color index from the 
Potsdam model fit described above and shown in Fig.~1.  
The slope of the spectral energy distribution is matched best for  
a reddening of $E_{B-V}$=1.45\,mag. For the correlation between  
color excess and hydrogen column density, we adopt the value given  
in Groenewegen \& Lamers (1989), $N_{\rm H}=(3.8\pm 
0.9)\times  10^{21}E_{B-V}~{\rm cm}^{-2}$. Then the interstellar column  
density of neutral hydrogen is 
$N_{\rm H}\approx 5.5\times 10^{21}{\rm cm}^{-2}$  towards WR\,114.  
 
\begin{figure}[hbtp] 
  \centering \mbox{\includegraphics[width=8cm]{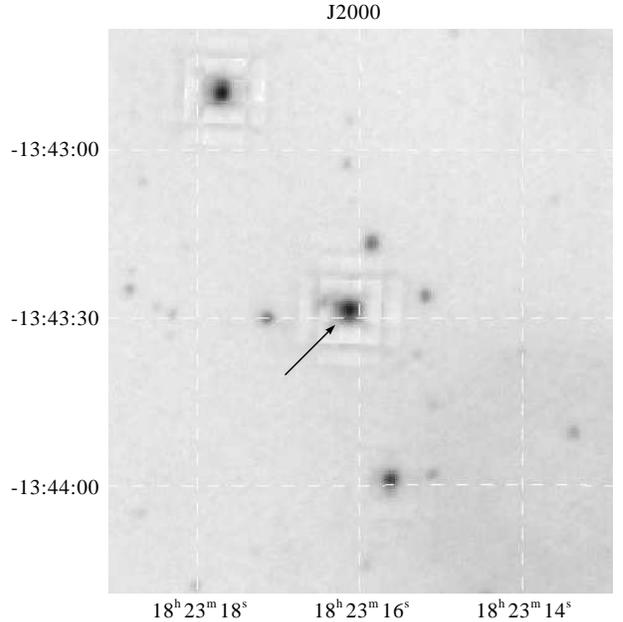}} \caption [ 
  ]{Low-resolution OM image in V~band of WR\,114. The star is clearly 
  seen in the center of the image as indicated by the arrow.} 
  \label{fig:f1} \end{figure}  
 
\begin{figure}[hbtp] 
  \centering \mbox{\includegraphics[width=8cm]{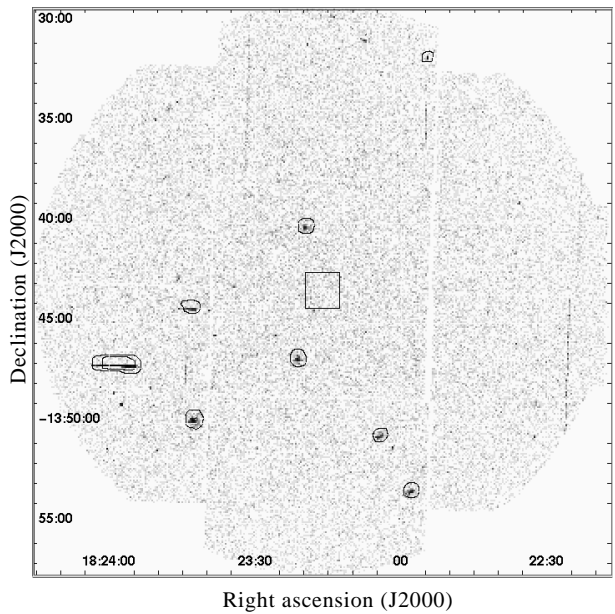}} \caption [ 
  ]{MOS1 X-ray image (0.2-10\,keV). The square box indicates 
   the same area of sky as shown in Fig.~\ref{fig:f1}.} \label{fig:f2} 
\end{figure}  
 
The integrated X-ray flux, $F_{\rm X}$, can be estimated by  
multiplying the observed count rate, CR, by an energy conversion factor  
ECF, that depends  
on the shape of the X-ray spectrum  
$F_{\rm X}={\rm ECF} \times {\rm CR}$. 
To infer an upper 
limit on X-ray flux for WR\,114, when there is no spectrum, 
we may only suggest some plausible model  
and treat the result with care. For OB stars, the ECF depends mainly 
on the X-ray temperature of  
the emitting plasma and the interstellar absorption  
(Bergh{\"o}fer \etal\, 1996). We assume a temperature of 
$k\Tx=$1\,keV. Then the count rate is converted into flux using  
ECFs provided by the \xmm\, Survey Science Center and the value of neutral 
hydrogen column density from our visual spectral fit.  
Since the distance to the star, $d=2$\,kpc, is known owing to the star's 
membership in the stellar association Sgr OB1 (van\,der\,Hucht 2001), 
we estimate an upper limit to the X-ray luminosity (0.2-10 keV) of 
\Lx\,$\lesssim\,2.5\times 10^{30}$\,erg\,s$^{-1}$, yielding 
$(\Lx/\Lbol)\lesssim\,4\times10^{-9}$.  It is necessary to emphasize that 
this upper limit is in fact the maximum \xmm\, sensitivity for a given 
exposure time and level of internal and cosmic background.   
 
\begin{table} 
\begin{center} 
\caption{Brightest X-ray sources detected by \xmm\, in the vicinity of 
WR\,114  
\label{tab:sources}} 
\begin{tabular}{l c c c }\hline\hline 
\rule[0mm]{0mm}{4.0mm} 
XMMU$^a$ &  Count Rates    &  Optical Counterpart  \\ 
         &  [\cps]         &    [mag] \\ \hline
J182342.8-134949 & $0.14\pm\,0.009$ & star, V=13.36 \\ 
J182321.2-134645 & $0.05\pm\,0.005$ & star, V=14.33 \\ 
J182258.1-135324 & $0.12\pm\,0.01~$ & -- \\ 
J182304.3-135038 & $0.06\pm\,0.005$ & -- \\ 
J182319.8-134011 & $0.04\pm\,0.005$ & -- \\  \hline \hline   
\end{tabular} 
\end{center} 
{\small $^a$ according to the naming convention posted on the 
\xmm\ website.} \\ 
\end{table} 
 
Other sources were detected during our abortive attempt on WR114 as clear  
from the MOS1~image shown in Fig.~\ref{fig:f2}. Some of  them are listed 
in Table~\ref{tab:sources}. The optical counterparts were identified using 
``The STScI Guide Star Catalog 2.2'' and the offsets 
between centers of X-ray images and optical images of the stars are $0\farcs4$ 
for XMMU J182342.8-134949 and $0\farcs1$ for XMMU J182321.2-134645. 
The latter source is 1.4\,arcmin from  
the pulsar PSR\,J1823-1347 which has no known optical counterpart although 
the much closer coincidence with a star of V=14.33\,mag is probably more 
relevant. It is worth noting that X-ray sources with optical stellar  
counterparts have little emission above 2\,keV and are probably galactic 
while those without optical counterparts have no X-rays below 2\,keV 
and are probably either young stellar objects or extragalactic. 
 
\section{X-ray observations of single WC~stars} 
 
There has been steady progress in the study of stellar X-ray emission.  
The cornerstone imaging X-ray observatories of the 20$^{th}$\,century --  
{\sc Einstein} (0.2-4.0~keV), {\sc Rosat} (0.2-2.4~keV),  
and {\sc Asca} (0.4-10~keV) -- have now finished and  
the new generation observatories \xmm\, (0.2-15~keV) and  
{\sc Chandra} (0.08-10~keV) are in operation. With the completion of 
source catalogues from the {\sc Einstein}, {\sc Rosat}, and {\sc Asca} 
missions, it is time to review the X-ray properties of WC~stars. 
 
In order to do so we searched all available catalogs of  
{\sc Einstein}, {\sc Rosat}, and {\sc Asca} missions. The normal practice  
for the inclusion of a source in a catalogue is a likelihood detection 
threshold of 10. Thus it is possible that real but weak detections might 
have been omitted. The {\sc Rosat} All-Sky Survey Faint Source Catalogue 
(RASS-FSC) is derived from the all-sky survey performed during the 
{\sc Rosat} mission. The sources with a detection likelihood  
of at least 7 and containing at least 6 source photons are included in this  
catalogue.  
 
In addition, we retrieved available archival data and performed our own data 
reduction with no detection threshold with the purpose of pinpointing even 
fainter sources. The results are discussed below.  
 
Two sub-samples of WC~stars from ``The {\sc vii}$^{th}$ Catalogue of  
Galactic WR Stars'' (van\,der\,Hucht 2001) were defined. The first group  
was formed by nominating as single stars those 
not designated as  
spectroscopic or colliding-wind binaries in the catalogue. We have thus  
counted the d.e.l. stars with the single stars for the assessment of  
WC~star X-ray emission. There are 56 such stars in the  
{\sc vii}$^{th}$ catalogue.  We note that amongst them, there are a few  
objects with the presence of dust as indicated by the letter ``d'' in  
spectral classification.  
Also, one of the stars from our sample, WR\,90 (WC7), is a confirmed source  
of nonthermal radio emission (Chapman \etal\, 1999). Therefore, it  
is most likely that other binary systems are to be found in this sub-sample 
as well as single stars. Nevertheless, none of these WC~stars appears  
in any of the {\sc Einstein}, {\sc Rosat} or {\sc Asca} source catalogues.  
It seems that the non-detection of WR\,114 by \xmm\, confirms the results  
of more than 20 years of attempts to observe X-rays from WC~stars but to 
a greater degree of sensitivity.  
 
The second sub-sample from the {\sc vii}$^{th}$ catalogue includes  
spectroscopic and colliding wind binaries. There are 31 such systems 
in the catalog. All X-ray sources included in {\sc Einstein}, {\sc Rosat}  
or {\sc Asca} source catalogues amongst WC~stars belong to this group.  
 
We will now briefly discuss the archival observations of WC-type~stars. The 
first quantitative information on the X-ray emission of hot massive stars 
was obtained by the {\sc Einstein} observatory. 
Seven single WC~stars were observed with the IPC aboard {\sc Einstein} and  
none detected as shown in Table~\ref{tab:einstein} taken from Pollock (1987b).
It is usual practice to apply a source detection threshold of $\lambda>10$,  
far above any of the WC~star statistics, which are quite typical of random  
background fluctuations. The exposure times achieved by {\sc Einstein} are  
quite significant, giving flux upper limits around 
$F_{\rm X}\lesssim 10^{-14}$ \,erg\,cm$^{-2}$\,s$^{-1}$. 
 
\begin{table} 
\begin{center} 
\caption{\label{tab:einstein} 
         {\sc Einstein} observations of WC-type stars showing  
         the log-likelihood detection statistic, $\lambda$, and  
         the maximum-likelihood count rate} 
\begin{tabular}{r c r c c}\hline\hline 
\rule[0mm]{0mm}{4.0mm} 
 WR & type & time  & $\lambda$ &  Count Rate        \\ 
    &      & [ksec]&           &  [$10^{-2}$\,\cps] \\ 
\hline  
5   & WC6  &  6.1  &  0.0      &    0.0             \\ 
17  & WC5  &  5.5  &  0.0      &    0.0             \\ 
57  & WC7  &  6.5  &  0.0      &    0.0             \\ 
111 & WC5  &  4.3  &  1.0      &    0.2             \\ 
135 & WC8  & 10.8  &  0.3      &    0.1             \\ 
144 & WC4  & 18.3  &  0.0      &    0.0             \\ 
154 & WC6  &  4.3  &  0.3      &    0.1             \\  
\hline\hline  
\end{tabular} 
\end{center} 
\end{table} 
 
The results were further confirmed by the {\sc Rosat} mission. 
The instruments, PSPC and HRI, aboard {\sc Rosat} were very occasionally  
used to observe WC stars during both the pointed programme and the All-Sky  
Survey (RASS). Although pointed observations had relatively long  
exposure times, the survey had the merit of observing essentially all 
Galactic WR~stars for a few hundred seconds.  
 
As pointed out by Pollock \etal\, (1995) in a compilation of {\sc Rosat} 
observations, single WR~stars are weak X-ray sources in general,  
with only upper limits for the majority. There is no single WC star which 
is included in any of the RASS catalogues.   
 
Since all of the data are now public we have extracted and analyzed the  
WC stars reported in Table~\ref{tab:ROSAT}. The table includes the visually  
brightest single WC~stars and those WC~stars which have available pointed 
observations. Once again, there are no detections to report. The apparent  
view in the literature that all types of WR~stars are sources of X-ray  
emission can only conceivably have been based on Pollock \etal\,'s (1995)  
report of a weak detection of WR\,111 using results from an early version  
of the {\sc Rosat} pipeline. With the more rigorous analysis reported here,  
we are unable to confirm even that feeble result. The flux upper limit is 
roughly $F_{\rm X}\lesssim 3\times\,10^{-14}$\,erg\,cm$^{-2}$\,s$^{-1}$. 
 
\begin{table} 
\begin{center} 
\caption{\label{tab:ROSAT} 
         {\sc Rosat} observations of prominent WC~stars$^a$}
\begin{tabular}{r c c l c c c } 
\hline\hline 
\rule[0mm]{0mm}{4.0mm} 
WR  & type &   V   & time & $\lambda$ & Count Rate        & \\  
    &      & [mag] &[ksec]&           & [$10^{-3}$\,\cps] & \\ 
\hline 
23  & WC6  & ~9.67 & ~2.8 &    0.7    & $0.5\pm 0.2$      & PSPC \\ 
52  & WC5  & ~9.86 & ~0.2 &    2.4    & $0.0\pm 3.0$      & PSPC \\ 
90  & WC7  & ~7.45 & ~0.3 &    0.0    & $0.0\pm 4.0$      & PSPC \\ 
106 & WC9  & 12.33 & 15.7 &    0.0    & $0.0\pm 0.5$      & PSPC \\ 
111 & WC5  & ~8.23 & 18.9 &    3.0    & $0.9\pm 0.4$      & PSPC \\ 
135 & WC8  & ~8.36 & 11.0 &    2.1    & $2.1\pm 1.7$      & PSPC \\ 
144 & WC4  & 15.49 & 81.3 &    0.0    & $0.0\pm 0.1$      & HRI  \\ 
154 & WC6  & 11.54 & ~0.4 &    0.3    & $4.8\pm 6.7$      & PSPC  \\  
\hline\hline 
\end{tabular} 
\end{center} 
{\small $^a$ {\sc Rosat} observations of prominent WC~stars from either the 
short exposures of the PSPC all-sky survey or longer pointed 
observations with either the PSPC or the HRI detectors. 
The observation time reported is that combined from the one or more 
available pointings, often at different off-axis angles. The count rate 
is the mean.} 
\end{table} 
 
These long-standing doubts over the emission from  single WC~stars can now be  
confronted with the advent of more sensitive observations from the newer  
generation of instruments. Accordingly, we have retrieved all available  
{\sc Asca}, \xmm\, and  {\sc Chandra} archival data. For all the stars 
discussed, if not pointed out otherwise, the interstellar neutral hydrogen  
column density was obtained in the same way as for WR\,114 (see above) using 
van\,der\,Hucht's (2001) values of $E_{b-v}$.  
 
{\bf WR\,111} (WC5). {\sc Asca} images of weak diffuse X-ray  
emission near the area of sky where WR\,111 is located are in the public  
domain. On 10 Oct. 1993 a 29\,ksec GIS exposure and a 23\,ksec SIS exposure  
were taken, with no detection of a point source anywhere in the vicinity of 
WR\,111. {\sc Asca} was capable of $5\sigma$ point source detections 
at flux limits as low as  $\sim 4\times10^{-14}$\,erg\,cm$^{-2}$\,s$^{-1}$ 
(Tanaka \etal\, 1994). Although this value can be accepted as an upper limit 
for non-detected sources, we decided to obtain more precise estimates making 
use of standard processed files retrieved from the archive. 
 
We obtained 1$\sigma$ upper limits on the count rate using the number 
of counts in the area  which corresponds to the PSF, and the  
exposure time corrected for vignetting as listed in Table~\ref{tab:asca}.   
As described in the {\sc Asca} data reduction guide, for GIS at energies  
of about 1\,keV, the FWHM point spread function  is about 3\,arcmin. 
For SIS the PSF is smaller, about 1.5\,arcmin. We used 1$\sigma$ limits  
for Gaussian statistics as tabulated in Gehrels (1986). 
Using stellar parameters for WR\,111 from Gr{\"a}fener \etal\, (2002), 
the upper limit to the X-ray flux, corrected for the interstellar absorption, 
is $F_{\rm X} \lesssim\,5.0\times 10^{-14}$\,erg\,cm$^{-2}$\,s$^{-1}$.  
Therefore, the upper limit on the X-ray luminosity of WR\,111 is  
$\log\Lx/\Lbol\,\lesssim\,-7.8$. 
 
{\bf WR\,121} (WC9d) The vicinity of WR\,121 was observed with {\sc 
Asca} on 23 Mar 1999 for a 53\,ksec exposure time with GIS and 50\,ksec 
with SIS. There is an extended source of diffuse X-ray emission in the 
central part of the image. The outer region of this diffuse source  
is superposed with the celestial coordinates of WR\,121.  Nevertheless, 
the signal from the diffuse source is very weak, and so we determine 
an upper limit of  
$F_{\rm X}\lesssim\,2\times 10^{-13}$\,erg\,cm$^{-2}$\,s$^{-1}$ for the  
unabsorbed flux. The stellar parameters of WR\,121 are not well known.  
Therefore we adopted a typical luminosity of WC9 type stars 
of $\log\Lbol/\Lsun=5.3$ resulting in $\log\Lx/\Lbol\lesssim\,-7.4$. 
 
\begin{table} 
\begin{center} 
\caption{{\sc Asca} observations of WC-type stars$^a$ \label{tab:asca}} 
\begin{tabular}{r c c c }\hline\hline 
\rule[0mm]{0mm}{4.0mm} 
WR & time            &  Count Rates       &  \Lx/\Lstar   \\ 
    & [ksec]         & [$10^{-3}$\,\cps]  &               \\ \hline  
    & G~~~S          &  G~~~~S            &               \\ \hline 
111 & 29~~23         & 2.3~~~~2.3         & $2\times\,10^{-8}$ \\ 
121 & 53~~50         & 1.7~~~~0.9         & $4\times\,10^{-8}$ \\ 
144 & 36~~33         & 6.2~~~~1.8         & $4\times\,10^{-8}$ \\ \hline\hline 
\end{tabular} 
\end{center} 
{\small $^a${\sc Asca} observations of WC-type stars showing 
1$\sigma$ upper limits to S, the average of SIS1 and SIS2, and G, the 
average of GIS1 and GIS2, and the corresponding luminosity.} 
\end{table} 
 
{\bf WR\,144}\,(WC4) This star was in the FoV of {\sc Asca} observations of 
the Cygnus\,OB2 stellar association.  During  a 36.3\,ksec {\sc GIS} 
exposure and 33\,ksec SIS exposure on 29 Apr. 1993 (Kitamoto \& Mukai 1996), 
the four point-like sources are the trapezium of O stars from Harnden 
\etal\,'s (1979) discovery paper. Within the 1$\sigma$ limit, there is no 
other point source in the image. There is however a  large amount of 
diffuse X-ray emission in Cyg~OB2 which spreads out to  the region where 
WR\,144 is located. Because WR\,144 is a highly reddened  
star, its stellar parameters are not well-known. Using the stellar parameters  
of WR\,111 and adopting $E_{B-V}=1.82$\,mag from Massey \etal\, (2001), we  
estimate $F_{\rm X}\lesssim\,1.5\times 10^{-13}$\,erg\,cm$^{-2}$\,s$^{-1}$, 
yielding $\log\Lx/\Lbol\lesssim\,-7.4$. 
 
{\bf WR\,60}\,(WC8) WR\,60 was in the field of view of the EPIC pn 
instrument on board \xmm\,  during observations of the Galactic Plane  
on 6-7 Feb. 2002. A total exposure of $\sim$12.7\,ksec was achieved.   
But since WR\,60 is located at $\sim$7.2\,arcmin off-axis, the useful  
exposure time is only 2.49\,ksec as listed in Table~6. There is no 
point-like source in this area in any of the energy bands. Due to the large 
offset from the center of the image, the sensitivity is much smaller than 
in the case of WR\,114. We estimate a 1$\sigma$ upper limit to the count rate  
of $1.5\times 10^{-3}$\,\cps. Then the unabsorbed flux is  
$F_{\rm X}\lesssim\,1.2\times 10^{-14}$\,erg\,cm$^{-2}$\,s$^{-1}$.  
The nominal luminosity of WC8 stars is $\log\Lbol/\Lsun\,=5.3$, giving  
$\log\Lx/\Lbol\lesssim\,-8.2$ for WR\,60. 
 
{\bf WR\,118}\,(WC9d) The star was in the FoV of \xmm\, on 
15 March 2002, with a useful exposure time of 3.8\,ksec for MOS and 
3.9\,ksec for pn. There is no point-like source detected in the  
X-ray image in the area where WR\,118 is located.  The 1$\sigma$  
upper limit to the count rate in the pn detector is $5\times10^{-4}$\,\cps.   
From this we infer an unabsorbed 
flux of $F_{\rm X}\lesssim\,4\times 10^{-14}$\,erg\,cm$^{-2}$\,s$^{-1}$, 
and adopting a typical luminosity for WC9~type stars, we obtain  
$\log\Lx/\Lbol\lesssim\,-7.4$.

{\bf WR\,101a-o, WR\,102a-l} There are six late-type WC~stars in the 
Sgr~A  region of the Galactic Center, which are highly reddened 
($E_{B-V}\approx\,9.5$\,mag).  Little can be said about these 
poorly studied stars. The properties and evolution of massive stars  
in this  region may be different (see e.g. Coker \etal\, 2002) 
due to the exceptional conditions in the vicinity  of the central Black Hole.  
The region has been extensively observed  by nearly all X-ray satellites  
including  two {\sc Chandra} exposures. We retrieved the archival 
{\sc Chandra} data and performed a crude analysis. None of the X-ray point 
sources coincide with optical images of any of the WC~stars in this region but 
the level of X-ray background is very high. Thus, no additional  
information in our quest on intrinsic X-ray emission from single WC~stars  
could be drawn from these extremely interesting objects.  
 
The same holds for the members of the Quintuplet cluster, one of  
the three massive clusters projected within 50 pc of the Galactic center  
(Figer \etal\, 1999), which contains at least 8 WC stars. The stars are 
highly reddened ($E_{B-V}\approx\,9.5$\,mag) and many of them are likely 
surrounded by dusty envelopes. It is therefore not surprising that 
no X-ray emission associated with the intrinsic properties of the 
stellar winds has been reported. 
 
\begin{table} 
\begin{center}
\caption{\label{tab:xmm} 
         \xmm\, observations of WC-type stars reporting 
         1$\sigma$ count-rate and luminosity upper limits}  
\begin{tabular}{r c c c c }\hline\hline 
\rule[0mm]{0mm}{4.0mm} 
 WR &   time    & Count Rates &  \Lx/\Lstar  \\  
    &   [ksec]  & [$10^{-3}$\,\cps] &     \\ \hline 
    & MOS~~~~pn & MOS~~~~pn    &                    \\  \hline    
60  & ---~~~~~2.5 & ---~~~3.3    &  $6\times10^{-9}$ \\ 
114 & 15.9~~~10.2 & 0.6~~~0.9    &  $7\times10^{-10}$ \\ 
118 & ~3.8~~~~~3.9 & 1.2~~~0.5    &  $4\times10^{-8}$  \\ \hline\hline  
\end{tabular} 
\end{center} 
\end{table} 
 
\section{X-ray observations of binary systems containing a WC~star} 
  
Extensive observational studies of WR stars in the Galaxy  
(see references in the  {\sc vii}$^{th}$ Catalogue) suggest that  
the fraction of binaries among WC stars is quite large, probably  
more than half. Thus, it is likely that  
many WC-type stars are not yet detected binary systems of (a) two massive  
stars, usually WC+OB~type, in which X-rays are generated by colliding stellar  
winds (e.g. Usov 1992), or rarely, (b) a WC~star and a compact  
companion, such as Cyg X-3, with much more luminous accretion-powered X-rays.
   
Such is probably the case of WR\,65 (WC9d). It is considered a binary in  
van\,der\,Hucht (2001) on the basis of the presence of diluted emission  
lines. We found that WR\,65 is over-luminous for its spectral type: 
from our  analysis of optical and UV spectra, we estimate 
$\Lbol=10^6\Lsun$. It is also  a radio source, with indications of being a  
non-thermal radio emitter (Chapman \etal\, 1999).  
 
\begin{figure}[hbtp] 
  \centering 
  \mbox{\includegraphics[width=8cm]{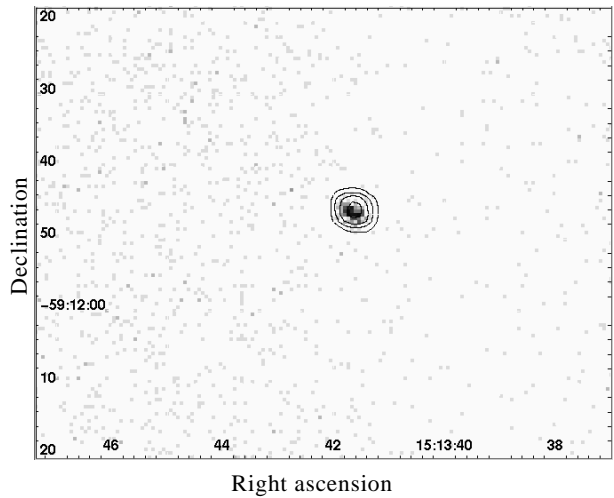}} 
\caption [ ]{Part of the {\sc Chandra} ACIS-I~image of the H{\sc ii} region  
in the vicinity of the pulsar PSR B1509-58.   
The point-like X-ray source is $1\farcs6$ 
from the optical position of WR\,65.} 
\label{fig:f6}  
\end{figure}

Since WR\,65 is 4~arcmin from one of the 
brightest X-ray pulsars in the Galaxy, PSR\,B1509-58,  
it has been in the FoV of nearly all X-ray satellites.  However, until the   
excellent {\sc Chandra} angular resolution became available,  
the star had not been resolved in X-rays. {\sc Chandra} observed  
the pulsar wind nebula on 14 August 2000 for a 20 ksec single exposure
 as reported  in Gaensler \etal\, (2002). Figure~\ref{fig:f6} shows 
part of the {\sc Chandra}'s ACIS-I image of the region in the vicinity 
of PSR\,B1509-58, showing the point-like source X-ray source coinciding 
with WR\,65. The neutral hydrogen column density derived by Gaensler 
\etal\, (2002) from fitting  the X-ray spectrum of this point-like  
source, $N_{\rm H}\sim 3\times10^{22}$\,cm$^{-2}$, is quite close to  
the value we derived from the reddening of the visual and UV spectra  
of WR\,65, with $E_{B-V}=2.42$ leading to $N_H\approx\,10^{22}$\,cm$^{-2}$. 
   
The {\sc Chandra ACIS-I} $5\sigma$ detection yields a count rate of 
$(8.87\pm\,0.8)\times10^{-3}$\cps, an unabsorbed flux of 
$F_{\rm X}\gtrsim 2\times 10^{-13}$\,cm$^{-2}$\,s$^{-1}$, and 
$\log\Lx/\Lbol\gtrsim\,-6.8$. However, there is  
some ambiguity in the identification of  
X-ray point source in the ACIS-I image as WR\,65. The X-ray source has  
an offset $\Delta\alpha=-0\farcs009$  and $\Delta\delta=+1\farcs63$  
from WR\,65 in FK5 (2000) coordinates. At the same time the radio coordinates  
of the source identified as WR\,65 in (Chapman \etal\, 1999) differ from the  
X-ray source in {\sc Chandra}'s FoV by only $\Delta\alpha=0\farcs003$  
and $\Delta\delta=-0\farcs13$. It seems likely, that the radio and  
X-ray sources are identical. Gaensler \etal\, (2002) 
reported an uncertainty of $\pm 0\farcs5$ in each coordinate for the  
pulsar's X-ray position.     
 
From broadband X-ray photometry it is already clear that WR\, 65  
is a rather hard source, with an emission maximum in the range 2-4\,keV  
and only trace emission in softer passbands  
(see Fig.~6 from Gaensler \etal\, 2002).  This is typical for the 
colliding wind binaries, such as $\gamma$~Velorum and WR\,140, but 
hard to explain in terms of a single star. Skinner \etal\, (2002a) has  
reported on a hard component ($kT\geq3$\,keV) in the spectrum  
of WR~110, a nitrogen-rich WN star that is believed to be single.   
However, in this case the dominant contribution to the X-ray emission  
is rather soft with a characteristic temperature k$T\approx0.5$\,keV. 
Therefore, combining data on the shape of X-ray spectrum with radio and  
bolometric luminosity excess and the presence of dust we conclude 
that WR\,65 is a newly discovered X-ray bright colliding wind system.   
It is clear that the nature of the X-ray emission of WR\,65 should be  
verified by further observations.   
 
Although the mechanisms of X-ray production in binary systems 
are thought to be  reasonably well understood, only 7 
WC~binary stars have been bright enough to be detected 
by one or more of the previous generations of X-ray instruments, namely  
WR~11, WR~48, WR~79, WR~93, WR~125, WR~137 and WR~140. 
All are spectroscopic binaries with orbital periods ranging between 
days and years. It looks as though we may now add WR~65 to their number. 
Seven WC~stars located in the Sgr~A and Quintuplet clusters have either  
ambiguous or false detections or large positional errors as indicated in  
object catalogues. 
Thus an overall fraction of detected X-ray sources is $\approx$10\% among all  
WC~stars. Even among the total number of the 37 binary WC~stars listed  
in the {\sc vii}$^{th}$ Catalogue, six deemed so only by virtue of diluted  
emission lines, about 20\% are detected X-ray sources. 
In this regard, the weakness of the X-ray emission from WR~114 has little 
or no bearing on its possible binary status.    
 
By far the brightest two X-ray WR binary systems, WR\,140 (O4-5{\sc v}+WC7)  
and $\gamma$~Velorum (WR\,11, O7.5{\sc iii}+WC8), both have WC~stars and  
have been observed reasonably often  (e.g. Williams \etal\, (1990), 
Stevens \etal\, (1996), Zhekov \& Skinner (2000), 
Skinner \etal\, (2001)). Both show variable absorption dependence on binary 
phase. WR140 is optically thin for most of its 7.94-year orbit, but shows 
strong absorption of X-rays near periastron. By contrast,  
absorption is heavy during most of $\gamma$~Velorum's 80-day orbit,  
reaching a still substantial minimum when the O-star is in front.  
\xmm\, observations of  $\gamma$~Vel have been reported by  
Dumm \etal\, (2002). At maximum light, the hard X-ray  
emission zone is clearly visible, though still absorbed below about  
2\,keV. At minimum light and throughout the orbit, a constant soft X-ray  
component is present that is convincingly explained by Dumm \etal\, (2002)  
as emission from cool,  
distant WC material photoionized by the central collision zone.  
In both systems, the WC star acts only as a passive X-ray absorber  
of dimensions of a few AU.  
 
Thus, we may conclude from the available  spectral and temporal data  
on binary systems, that there is little or no evidence for WC-type  
stars being intrinsic X-ray sources, and that WC~stars are effectively  
opaque to X-rays from the colliding wind zone. The small percentage  
of WC binary X-ray sources shows that the conditions under which such a  
system can emit X-rays are rather special, requiring particular combinations  
of orbital dimensions and viewing geometry.

  
\section{Discussion}  
  
New spectral windows, larger and better telescopes, and advances   
in technology have led and continue to lead to important new discoveries   
about well-known astrophysical sources, and even whole new classes of   
sources. Such has also been the case with X-rays from hot massive stars.  
  
Although there is no doubt that OB stars emit X-rays, neither  
the properties of the ensemble or individual spectra are   
well explained. The rough scaling relationship between bolometric   
and X-ray luminosity of OB~stars $\Lx\propto 10^{-7}\Lbol$ established  
with early {\sc Einstein} observations by Pallavicini \etal\, (1981) 
has been interpreted as a consequence of the coupling of the wind  
momentum-luminosity relation of Kudritzki \etal\, (1999) with a special  
distribution of hot  plasma in the wind  (Owocki \& Cohen 1999). 
The X-ray emitting WN~stars do not follow  this scaling  
law (Wessolowski 1996) although Ignace \& Oskinova (1999) 
showed that the measurements could be understood if the X-ray plasma 
filling factor depends on mass loss rate and terminal velocity.

Whereas {\sc Einstein, Rosat} and {\sc Asca} could provide only gross 
energy distributions, {\sc Chandra} and \xmm\, have been able to 
resolve individual lines and thus challenge further the 
current view of X-ray production in radiatively-driven winds. The 
theoretical X-ray spectra of hot stars with wind-distributed shocks are 
expected to produce asymmetric emission line profiles with blueshifted 
emission peaks owing to wind absorption effects (e.g. Ignace 2001; 
Owocki \& Cohen 2001; Ignace \& Gayley 2002).  Recent results 
from {\sc Chandra} and \xmm\, have shown that most of the OB~stars 
observed so far display symmetric and unshifted emission profiles (e.g. 
Schulz \etal\, 2000, Waldron \& Cassinelli\, 2001).  Although the 
O~star $\zeta$~Pup is a notable exception, with profiles better 
matching theoretical expectations (Kahn \etal\, 2001, Cassinelli 
\etal\, 2001), it has not yet been possible to explain symmetric line 
profiles from single stars in the framework of the shock model. 
 
The WR~stars are another class of object in which the X-ray emission is 
thought to arise from a similar mechanism.  The WR~stars, being 
fainter, are an integral step behind the OB~stars in terms of spectral 
resolution.  {\sc Rosat} and {\sc Asca} provided passband fluxes for 
WR~stars, and now \xmm\, is providing spectral shapes. The 
characteristic feature of embedded shock models is the soft X-ray 
spectrum, with maximum temperatures corresponding to a fraction of 
terminal velocity of the wind. The emission is not expected to be present 
at energies harder than $\sim$2-3\,keV. However, Skinner \etal\, (2002a,b) 
have reported ``hard tails'' (i.e., the presence of emission with 
$k\Tx\geq3$\,keV) in the spectra of the WN stars WR\,6 and WR\,110.   
We have observed WR\,1, which is of similar subtype as these other  
two WN stars. WR\,1 does not appear to have a hard tail; 
however, the exposure for this source was less than for WR\,6 and WR\,110, 
so that the existence of a hard tail in WR\,1 is inconclusive at this 
point (Ignace \etal\, in prep). The 
hard tail may indicate the presence of non-degenerate companions which 
have otherwise gone undetected in these systems, an option that is also 
discussed in an attempt to explain the X-ray line profiles in 
O~stars. 
 
\begin{figure}[hbtp] 
  \centering \mbox{\includegraphics[width=8cm]{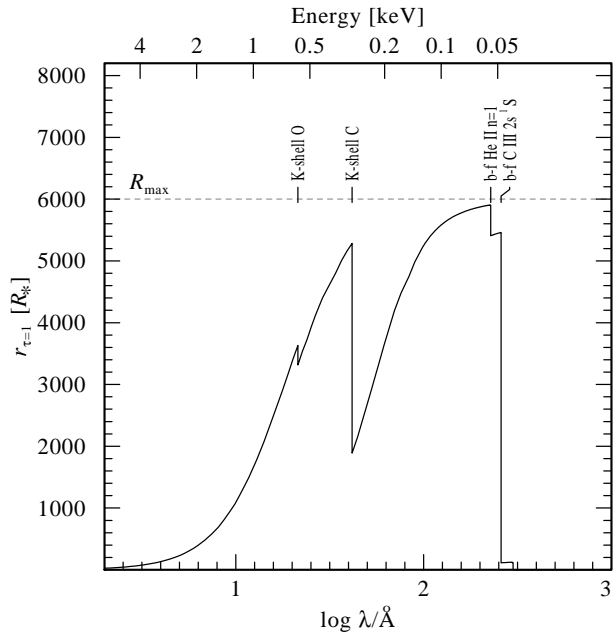}} \caption 
  [ ]{Radius where the photoabsorbing optical depth is unity 
to X-rays for the WC~star WR\,114 as obtained from the Potsdam atmosphere 
code. The model parameters are as given in \S~2.  
The lower horizontal axis is for wavelength, and the upper is for 
energy. The 
most prominent He\,{\sc ii} edge at 228\,\AA\ remains optically thick to 
the outer boundary of our model at $R_{\rm max}$ = 6000\,\Rstar.} 
  \label{fig:f8} \end{figure}  
 
In contrast to OB and WN stars, this paper reports on the effective absence  
of X-ray emission from WC~stars, another unexpected result. The obvious  
question is, why are these stars so faint in X-rays?  There are two obvious 
alternatives :  
 
(1) WC~stars are intrinsically inefficient in producing X-rays.  If 
      we understood the production of X-rays by radiatively-driven winds 
      better,  it would be easier to assess this possibility. Our  
      understanding is  hampered by the fact that  time-dependent 
      hydrodynamical simulations  including multiline scattering -- the latter 
      being central both for the  wind driving and instability growth rates 
      -- are not available yet. 
 
(2) X-rays are generated in the stellar winds of WC~stars  
      (as it is the case in OB star winds) but 
      the strongly enhanced density and metal content of their 
      winds lead to their total absorption before 
      they can emerge to the observer. 
 
The high optical depth for X-rays in WR~winds has long been recognized 
(Pollock 1987a), and it is this possibility that seems the most 
likely. As an O~star evolves through the WN~stage,  it converts 
4H$\rightarrow$He with an enhancement of N as the wind density 
increases. Both factors independently cause the X-ray opacity to 
rise. This is especially pronounced in cases where helium 
recombines to He{\sc ii}, primarily in the winds of late type WN~stars. 
The He{\sc ii} edge for those 
stars is strong enough to place the $\tau=1$ radius to several hundred 
stellar radii for the whole soft X-ray band. 
 
For the WC~stars, the abundance of CNO elements is drastically increased  
(while abundances of heavier elements like S, Si and Fe are unchanged) and 
the mass loss rates are the same order as for WN~stars.  It is 
clear that such a metal-rich and dense medium is very efficient in absorbing 
X-rays, making WC~winds even more opaque to X-rays than the winds of O and 
WN~stars. 
 
We have checked this by means of our advanced non-LTE atmosphere model, 
accounting for line blanketing and clumping in a first approximation. 
Although the latter effect leads to smaller empirical mass loss  rates 
(Hamann \& Koesterke 1998)  compared with those found previously from 
homogeneous models, the radius of optical depth unity is still 
extremely large. As seen in Fig.~\ref{fig:f8}, although He{\sc ii} remains 
the major absorbing element for the soft X-rays, the high abundances of CNO 
group elements leads to  strong K-shell absorption edges and  places 
the X-ray photosphere at distances exceeding a thousand stellar 
radii for photons with energies around 1\,keV. 
 
The upper limit of X-ray luminosity for WR~114 derived in this  
paper refers to the {\it emergent} stellar radiation after correcting 
for interstellar absorption. We cannot infer anything 
about the possible generation of X-rays by any hot plasma  
embedded deep in the wind that would be absorbed before they can escape. 
There is much room for such internal X-ray absorption as the optical  
depths in the wind of WR~114 is high to large distances.   
Without any knowledge about the location of the possible X-ray  
production and the detailed structure of the absorbing cool wind,  
quantitative limits on the internal X-ray production rate would be  
rather speculative. 
 
It would be extremely difficult for the theory of instabilities in  
radiatively driven winds to explain the presence of X-ray emitting 
material at distances of a thousand stellar radii. There are few 
detailed physical studies of the X-ray emission produced at such large radii. 
Feldmeier \etal\, (1997a,b) performed hydrodynamical simulations of 
O~type star winds for radii up to  100\,\Rstar. The hot X-ray emitting 
gas can be generated by shocks only in the accelerating part of the 
stellar wind, which does not extend further than 5--10\,\Rstar. 
When the expanding material reaches the constant velocity regime there 
are still regions of hot gas confined between dense and cold clumps. 
Nevertheless, the density of the interclump medium is 2-3 orders of 
magnitude lower than the density of clumps, hence cooling is not 
efficient and no X-rays  can be generated.  Runacres \& Owocki (2002) 
confirmed the major  conclusions of Feldmeier \etal\, (1997a,b) and 
pointed out that the clumpy structure of the wind persists to outer 
regions with radii up to 100\,\Rstar\,  and is maintained by collisions 
between clumps moving with different speeds, at least in a 1-D approach. 
Clump-clump collisions in the outer wind could hardly be a means for 
generating X-ray emitting plasma, since the relative velocities of the 
clumps are most likely not high enough. The slight chance that  slowly 
approaching clumps can squeeze the thin hot gas between them to reach 
the densities relevant for X-ray emission should be explored via 
hydrodynamical  simulations, but likely it cannot serve as the main 
mechanism for the hot plasma production. 
 
There is no study known to us which would trace the evolution of wind 
structures as far out as the X-ray photosphere of WC~stars. In any case,  
from the discussion above it seems implausible that X-rays could be 
generated so far out in the stellar wind. Thus, although  
from this point-of-view, the absence of X-rays emerging from WC  
stars is consistent with the theory of radiatively driven winds,  
the wealth of new high-resolution spectroscopic data  
will undoubtedly stimulate future theoretical developments  
of the X-ray production in hot star winds. 
 
 
\section{Conclusions} 
 
We have analyzed the available X-ray archival data of WC~stars and 
a 15.9\,ksec \xmm\, observation of WR\,114. The main points are: 
 
({\sc i}) No single WC~star has previously been detected as an 
X-ray source, independent of spectral sub-type or other stellar 
parameters. 
 
({\sc ii}) \xmm\, observation of the typical WR~star WR\,114 (WC5) 
confirmed that this carbon enriched Wolf-Rayet star does not emit X-rays
above the detection limit of our observation. The upper limit to the 
X-ray flux from WR\,114 is $F_{\rm X}\lesssim 5\times 
10^{-15}$\,erg\,cm$^{-2}$\,s$^{-1}$. 
 
({\sc iii}) The apparent absence of X-rays can be explained by the 
large absorptive opacity of the stellar wind from WC~stars: the X-ray 
photosphere is at a few thousand stellar radii. 
 
({\sc iv}) All WC~stars that are currently known to emit X-rays  
are binary systems. The fraction of X-ray sources among all binary systems  
containing a WC~star is less than 20\%, indicating that the conditions 
are special under which they are visible.    
 
({\sc v}) A previously unidentified source of X-ray emission is attributed 
to the WC9d~type star WR\,65. Because of the hardness of its 
X-rays, we suggest that it too is a binary system. 
 
\begin{acknowledgements} 
 
We acknowledge useful discussions with  A.~Feldmeier and G.~Gr{\" a}fener.  
We wish to thank I.~Antokhin for the guidance on using \xmm\,'s SAS.  
This research has made use of NASA's Astrophysics Data System Bibliographic  
Services and of the SIMBAD database, operated at CDS, Strasbourg, France.  
Data were obtained through  the High Energy Astrophysics  
Science Archive Research Online Service, provided by the NASA/Goddard  
Space Flight Center. LMO acknowledges support for this research from the  
Deutsche Forschungsgemeinschaft grant Fe 573/1-1.  RI acknowledges  
support for this research from NASA grant NAG5-12557. RI and JCB would like  
to thank PPARC for the support and LMO is grateful for the NATO/RS fellowship.  
 
\end{acknowledgements}

\end{document}